# Resilience Dynamics of Urban Water Security and Potential of Tipping Points

E. H. Krueger[1,2], D. Borchardt[1], J. W. Jawitz[3], H. Klammler[4,5], S. Yang[2], J. Zischg[6] and P.S.C Rao[2,7]

[1]Department of Aquatic Ecosystem Analysis, Helmholtz Centre for Environmental Research - UFZ, Leipzig, Germany
[2]Lyles School of Civil Engineering, Purdue University, West Lafayette, Indiana, USA
[3]Soil and Water Sciences Department, University of Florida, Gainesville, Florida, USA
[4]Engineering School of Sustainable Infrastructure and Environment (ESSIE), University of Florida, Gainesville, Florida, USA
[5]Department of Geosciences, Federal University of Bahia, Salvador, Bahia, Brazil
[6]Unit of Environmental Engineering, Department for Infrastructure, University of Innsbruck, Innsbruck, Austria
[7] Department of Agronomy, Purdue University, West Lafayette, Indiana, USA

Corresponding author: Elisabeth Krueger (elisabeth.krueger@ufz.de)

**Key Points:**

- Data-model analyses of seven global cities reveal coevolution of security and resilience in urban water supply systems.

- Global and climate change with increasing, unpredictable shocks can push urban water systems across tipping points.

- Resilience emerging from community adaptation is highly variable across and within cities, leading to high inequality and precariousness.

**Abstract**

Cities are the drivers of socio-economic innovation, and are also forced to address the accelerating risk of failure in providing essential services such as water supply today and in the future. Here, we investigate the resilience of urban water supply security, which is defined in terms of the services that citizens receive. The resilience of services is determined by the availability and robustness of critical system elements, or "capitals" (water resources, infrastructure, finances, management efficacy and community adaptation). We translate quantitative information about this portfolio of capitals from seven contrasting cities on four continents into parameters of a coupled systems dynamics model. Water services are disrupted by recurring stochastic shocks, and we simulate the dynamics of impact and recovery cycles. Resilience emerges under various constraints, expressed in terms of each city's capital portfolio. Systematic assessment of the parameter space produces the urban water resilience landscape, and we determine the position of each city along a continuous gradient from water insecure and non-resilient to secure and resilient systems. In several cities stochastic disturbance regimes challenge steady-state conditions and drive system collapse. While water insecure and non-resilient cities risk being pushed into a poverty trap, cities which have developed excess capitals risk being trapped in rigidity and crossing a tipping point from high to low services and collapse. Where public services are insufficient, community adaptation improves water security and resilience to varying degrees. Our results highlight the need for resilience thinking in the governance of urban water systems under global change pressures.

**Plain Language Summary**

We evaluated the resilience of global urban water supply systems, including in economically advanced cities and those characterized by a prevalence of informal settlements lacking basic infrastructure services. Global Change challenges urban resilience with more frequent floods, droughts, population growth and competition for resources. We demonstrate that urban resilience requires the availability of financial and other "capitals" (water, infrastructure, efficient governance institutions) as well as robust responses to extreme events. Application of a systems dynamics model shows the impact of and recovery from repeated shocks for each city, and that tipping points may be crossed, if changing conditions are not adequately addressed. When public water supply services fail, citizens adapt by buying water from private vendors or storing water in rooftop tanks. However, inequality in adaptive capacity exists within and across cities, and leaves citizens in precarious situations. This impairs development potential and cities risk getting caught in a poverty trap. Focus on the development of "capitals", but neglect of robustness risks cities being pushed into a "rigidity trap" characterized by degrading services. Finally, predictions of the timing and magnitude of extreme events will remain unreliable, and managing for resilience will require managers to embrace probabilistic scenarios and uncertain futures.

# 1  Introduction

Common assessments of urban water supply security quantify the average per capita water availability (Damkjaer & Taylor, 2017; Floerke et al., 2018; Jenerette & Larsen, 2006; McDonald et al., 2011, 2014; Padowski & Jawitz, 2012), or focus on the sections of urban society living in water poverty (Cho et al., 2010; Eakin et al., 2016; Juran et al., 2017; Srinivasan et al., 2010; Sullivan, 2002; Wutich et al., 2017). Padowski et al. (2016) suggest that urban water security results from a combination of local hydrological conditions and management institutions in place that are capable of developing infrastructure for accessing regional water resources, as needed. Of 108 investigated cities in Africa and the US, 7% remain insecure due to minimal ability to access local and/or imported water. Floerke et al. (2018) produce water security scenarios for 482 of the largest cities worldwide by including the impacts of competitive uses among different sectors. Their results indicate that, by 2050, 46% of cities will be facing water security issues resulting from either surface water deficits or due to competitive conflicts with agricultural water use. In a review on water scarcity more generally, Liu et al. (2017) show that estimates of water scarcity lie in the range of 3-4.5 billion people affected in 2050. In addition to city-scale water insecurity, inadequate provision of available water resources within cities affects a much larger (as yet unquantified) portion of global urbanites. At least the 23 % of the total urban population worldwide living in informal urban settlements without adequate access to public urban services (883 million people) (UN, 2018) can be assumed being affected by water insecurity to a significant degree.

Integrated approaches able to capture water insecurity more broadly have largely been qualitative, theoretical or based on individual case studies (Eakin et al., 2017; Hale et al., 2015; Patricia Romero-Lankao & Gnatz, 2016; Wutich et al., 2017) (for reviews see Garfin et al., 2016; Garrick & Hall, 2014; Hoekstra et al., 2018). Recent research has emphasized the need for systematic approaches and metrics that are transferrable and scalable to allow cross-site comparisons, as well as those that combine quantitative metrics with context and qualitative information (Garfin et al., 2016; Wilder, 2016). These articles have also investigated the linkages between water security and adaptive capacity, embracing the concept of "capitals" (also referred to as "assets", "resources' and "desirable determinants" (DfID, 1999; Smit et al., 2001)) needed for adaptive capacity (Kirchhoff et al., 2016; Lemos et al., 2016; Varady et al., 2016). In a recent paper, Krueger et al. (2019) presented a quantitative, empirically-based and comparative method, which systematically integrates several forms of "capital" for adaptive capacity and provides context to highlight place-based nuances. The method was used to estimate urban water supply security in terms of the actual services that citizens receive, including access, safety (of access and water quality), reliability, continuity and affordability. In the presented framework ("Capital Portfolio Approach, CPA") the authors proposed that such services require not only the availability of water resources at the city level, but also the intra-urban infrastructure for storing, treating and distributing the water, financial capital and governance institutions and, when public services fail, community adaptation to cope with and adapt to insufficient water supply services. The analyses showed that, in cities with high levels of public services, community adaptation remains inactive as long as services perform as demanded. Cities with high levels of water insecurity rely on community adaptation for self-provision of services. Therefore, variability in urban water security is highly dependent on community adaptive capacity. We adopt this notion of water supply security, and assess its resilience in response to recurring shocks and disturbances.

The literature on adaptive capacity and water security has revealed tensions in the assessment of water security in static versus dynamic terms (Lemos et al., 2016). Water security approaches, which describe the state of a system at a certain point in time fall short of capturing the dynamic drivers and adaptive response of human actors suffering from water insecurity. There is also a need for capturing the adaptive response of the human actors and their ability of changing the system. Pathways approaches are suggested to examine the long-term evolution and path-dependent trajectories of water security (Lemos et al., 2016), where risk-based approaches consider the potential changes emerging in an uncertain future (Garrick & Hall, 2014).

One way to address such dynamics is through resilience approaches. Resilience of urban water supply services refers to the dynamic behavior of the system in response to disturbances; its ability to absorb shocks, its adaptation and reorganization in order to maintain system functions (here: water supply services) (Gunderson & Holling, 2002; Walker et al., 2004). The resilience of urban water systems is threatened by increasing demand driven by population growth, urbanization and life-style changes, as well as by changing land use and climate conditions (Floerke et al., 2018; Patricia Romero-Lankao & Gnatz, 2016). Thus, resilience is an emergent behavior in response to disturbances contingent on the timing and magnitude of shocks and requiring constant adaptive management (Allan & Bryant, 2011; Klammler et al., 2018; Park et al., 2013). Scheffer et al. (2018) discuss how the resilience of complex adaptive systems emerges from the interplay of several networked subsystems, where systemic resilience and the resiliencies of the sub-components are coupled in a non-linear way, with the possibility of tipping points. The need for the development of models and integrated systems modeling was recognized for the multiple domains of complex, interconnected and interdependent infrastructure systems (Linkov et al., 2014).

Resilient behavior, too, has been linked to adaptive capacity, which builds on the availability of a range of resources and assets (Brown & Westaway, 2011; Bryan et al., 2015; Eakin et al., 2014; Gallopín, 2006; Milman & Short, 2008; Waters & Adger, 2017). Such integrated approaches highlight the role of social actors, not only for human resilience, but also for the resilience of infrastructure systems (Rao et al., 2017). Waters and Adger (2017) present a fine-grained analysis of the relationship between adaptive capacity and resilience of urban informal settlement dwellers with emphasis on the heterogeneity of resilience in space and time based on a range of factors that include determinants of social capital, as well as urban form. While urban resilience at the city scale has been the focus of a number of studies and well-established index methods, they remain static or focused on resilience in general terms (Meerow et al., 2016; Spiller et al., 2015; UN-Habitat, 2017; UNISDR, 2017). We contribute to this literature by quantifying city-scale resilience of urban water supply security, including the heterogeneous role of community adaptation across cities.

Several models have been proposed that capture the dynamics of social-ecological, socio-technical and coupled natural-human-engineered (CNHE) systems. Some address specific systems and management responses, for example, reservoir management during flood and drought periods (Baldassarre et al., 2017). Emergence of poverty traps is observed in a CNHE systems model that investigates the dynamics of water security-related investment and its interplay with economic growth and risk reduction (Dadson et al., 2017). Muneepeerakul and Anderies (2017) introduce a model that shows how the coupled dynamics of the natural environment, infrastructure, public providers and resource users driven by financial incentives, give rise to the emergence of a governance system. Carpenter and Brock (2008) introduce a generic social-ecological systems model, where low, medium, and high levels of control ("stress") are associated with poverty traps, adaptive capacity, or rigidity traps. The system is forced externally through unexpected shocks, and recovery depends on adaptive capacity. With a similar logic, Klammler et al. (2018) introduced a CNHE systems model representing two state variables 1) critical infrastructure service deficit, and 2) adaptive management mobilized to recover services. They show how different model parameterizations can lead to multiple (stable) system states, and how sequences of recurring, stochastic shocks can lead to regime shifts from one state to another, or force the system into collapse. These existing models provide important insights into the dynamics and resilience of coupled systems. However, they remain theoretical and generic, without empirical application.

Here, we address these gaps by integrating system state, risk and dynamic response using empirical data. We consider the security of urban water supply as a system state, which is subject to shocks resulting from risks potentially threatening the system state. Resilience of the system refers to its dynamic behavior in response to shocks. The dynamic response requires adaptive capacity (based on a "stock of capitals"), and the action taken by human actors is the adaptive management marshaled by mobilizing adaptive capacity through capital robustness. We assess the resilience of urban water supply services for seven urban case studies located in contrasting hydro-climatic regions and a wide range of socio-economic conditions. The model framework of Klammler et al. (2018) is parameterized based on the Capital Portfolio Approach (CPA) of Krueger et al. (2019). Results by the former showed the existence of multiple stable states in the interaction of infrastructure services and adaptive management, as well as contingency of services on stochastic shock regimes. Results of the CPA analysis by the latter authors showed that capital availability and robustness are positively correlated, while risk correlates negatively. In combining the two, our research is guided by the hypotheses that: 1) The relationship between urban water security and resilience is non-linear, with potential for tipping points. If confirmed for urban infrastructure services, the non-linearity and tipping-point behavior found in natural and other coupled complex systems (Lade et al., 2013; Park & Rao, 2014; Scheffer, 2009) would make an interesting comparison. 2) Community adaptation increases city-scale water security and/or resilience to varying degrees due to different constraints in building security and/or robustness.

## 2 Methods

### 2.1 Systems Dynamics Model for Urban Water Supply Services

The model describes the temporal dynamics of two coupled system states, *service deficit* ($0 \leq \Delta(t) \leq 1$) representing the deficit of water supply services at the citizen scale and 2) *service management* ($0 \leq M(t) \leq 1$) representing service maintenance and recovery. Service deficit ($\Delta$) and service management ($M$) are aggregated (averaged) values for the entire system. The equations are derived from models describing the behavior of complex ecological systems and are applied here to complex urban systems. We use the scaled parameters and normalized equations, so that the model is *non-dimensional* (normalized to unit replenishment rate in adaptive capacity). The coupled temporal dynamics of these (dimensionless) state variables is written as (Klammler et al., 2018):

$$\frac{d\Delta}{dt} = (1-\Delta)b - aM\Delta + \xi \qquad (1)$$

$$\frac{dM}{dt} = (1 - c_1\Delta)M(1-M) - r\frac{M^n}{\beta^n + M^n} - c_2\xi \qquad (2)$$

where the first term on the right side of Eq. 1 represents growth in service deficit ($\Delta$ = 1-Service/Demand), which is the sum of demand growth and service degradation (rate constant *b*). The second term in Eq. 1 represents service recovery provided by M with efficiency coefficient (constant *a*). Stochastic shocks ($\xi$) lead to increases in service deficit, and are modeled as outcomes of a Poisson process, with mean frequency ($\lambda$) and exponentially distributed magnitude of mean value ($\alpha$). Replenishment in the capacity of service management (M, Eq. 2) follows a logistic function and is limited by coupling with $\Delta$ through $c_1$. For $c_1 => 0$ the two systems are increasingly decoupled. Capacity of M can be lost as a result of insufficient management efficacy and lack of financial capital, and can be aggravated by degrading infrastructure and insufficient water resources. The degradation of M follows a Langmuir (or Hill-type) function (Langmuir, 1918), determined by the maximum relative depletion rate (*r*). The shape of the depletion curve in M is characterized by the scale $\beta$ at which the degradation of service management begins to level off, and the shape (steepness) *n* of the degradation curve. Shocks can directly impact M, depending on the value of coupling parameter $c_2$.

Numerical simulations of time series are generated for the two state variables ($\Delta$ and M) by simultaneously solving Eq. 1 and 2 using a MATLAB ordinary differential equation solver (ode45), applied separately to each time interval between shocks. Shock magnitudes ($\alpha$) are added to $\Delta$ and subtracted from M at the end of each interval to form the initial value for the subsequent interval. Simulations are conducted for 1000 time units for each system, long enough with respect to mean shock arrival times and recovery time scales, such that states contained in a single realization are representative of average system behavior, and account for memory effects resulting from recurring shock impacts.

Dimensionless time (*t*) is scaled proportional to unit replenishment rate in service management ($t = t_{real} r_{RF}$; see (Klammler et al., 2018) for further details on normalization of Eq. 1 and 2). This means that *t* represents varying lengths of real time, depending on each type of city and the magnitude of shock impacts, and can be in the order of days or weeks (or less for resilient and water secure cities), or months to years (or even decades for non-resilient and transitional cities). While there are no comprehensive long-term empirical data on recovery times in response to different types of chronic and acute shocks in different cities, examples demonstrate that recovery is slower in poorer as compared to richer areas (Cutter & Emrich, 2015) (i.e., low versus high capital availability and robustness). A typical example of a chronic shock is supply intermittence due to the bursting of a water distribution pipe. Our data for Amman, Jordan, suggest that recovery from such shocks is in the order of ≤ 1 day. However, recovery from larger, less-frequent shocks may take much longer. For example, the 2017 Central Mexico earthquake disconnected 6 million people from the water pipe network. Most services were recovered within around two weeks (Audefroy, 2018), however recovery was an on-going process six months later (Hares, 2018; Unicef, 2018). Recovery from even more severe shocks, such as civil war, is a multi-year process. Liberia's capital Monrovia has faced severe water supply insecurity since the city's hydro-power plant - necessary for powering the water treatment plant and distribution system - was destroyed in 1990, at the beginning of the 14-year civil war (Smith et al., 2013). Even since the end of the war, water supply was secured only for 25% of the city's population, with improved prospects since the reconstruction of the hydro-power plant in 2016 (FPA, 2016a, 2016b). The total simulated time series of 1000 time units is therefore in the order of years to decades.

## 2.2 Capital Portfolio Approach

The CPA proposed by Krueger et al. (2019) considers 1) public services provided by a municipal entity and 2) total services resulting from a combination of public services and community adaptation in response to insufficient services (additional water bought on the private market, water stored and treated at the household level, etc.). Public services require four types of "capital": water resources (W, "natural capital", including naturally available and captured water resources), infrastructure (I, "physical capital") needed to store, treat and distribute W, financial capital (F) to build, operate and maintain the water supply system, management efficacy (P, "political capital") to operate and maintain services. Community adaptation ("social capital", A) complements or replaces insufficient public services. Three dimensions of these capitals are considered: availability, robustness and risks.

Two aggregate metrics represent capital availability required for public water supply services ($CP_{public}$={W, I, F, P}) and total services ($CP_{total}$=($CP_{public}$+A)), which includes the adaptation and additional "self-services" of the community. Robustness of public and total services ($RP_{public}$= *{$W_R$, $I_R$, $P_R$, $F_R$}* and $RP_{total}$={*$W_R$, $I_R$, $P_R$, $F_R$, $A_R$*}) and acute and chronic risk of shocks represent additional dimensions used for the parameterization of the model. Availability, robustness and risk are determined for each of the five capitals using scored and aggregated attributes, which are compiled across the five capitals in a hierarchical aggregation procedure using additive, and mixed additive and multiplicative aggregation methods (Krueger et al., 2019). An overview of adequate aggregation methods is provided in Langhans et al. (2014). Krueger et al. (2019) pay close attention to aggregation in terms of substitutability or multiplicative effects (e.g., one-out, all-out effects), but refrain from weighting the different metrics and sub-metrics in the CPA. While certainly expert weighting, such as proposed by several authors (Eakin & Bojórquez-Tapia, 2008; Patricia Romero-Lankao & Gnatz, 2016; Vincent, 2007) would provide more nuanced results, keeping the same relative metric weights (un-weighted) makes sense for this analysis for several reasons: 1) The specific objective function analyzed here (urban water supply security) aggregated at the city scale is the same across all case studies, and for the objective function to be achieved, the same set of capitals (with sub-metrics) is required for fully functional services (Krueger et al., 2019). If and how the diverse sub-metrics interplay in providing services is – to date – unknown. 2) Data availability is highly variable across case studies, and data uncertainty is high. Adding weights to (differentially) uncertain data would complicate and potentially distort the overall picture. Thus, refinement of the analyses for individual cities should be done in a co-production process with local stakeholders involved in expert judgement for assessment and potential weighting of sub-metrics. 3) The resilience analysis proposed here serves the purpose of understanding system dynamics and aggregated services as a fraction of total, rather than deciphering processes and interactions taking place inside the system. Weighting the different sub-metrics would be useful for understanding the interplay among the capitals (and sub-metrics), which is beyond the scope of the research presented here.

As laid out in the CPA, our investigations here explicitly address water supply services in terms of quantity. Water quality is implicitly addressed in several ways: 1) The ability of the public supplier to provide water at safely drinkable quality; 2) The quantification of community adaptation considers the need for treating water to make it drinkable; 3) The robustness of water resources considers the governance of water quality in a ranked scoring system; 4) Considered risks include contamination through dilapidated or lack of infrastructure (e.g., epidemic incidences caused by intrusion of sewage), as well as through industrial spills caused by upstream industry.

In addition, the CPA also implicitly addresses spatial and temporal dimensions. Access (spatial) to water services is considered in the quantification of the state of infrastructure through the household connection rate. The temporal dimension is considered in the continuity of supply through the need for the community to bridge temporal supply gaps (e.g., in rationed supply schedules).

We use these metrics representing the three dimensions of the CPA to reframe the model parameters. Resilience of water supply services is assessed by simulating impact-recovery cycles for the seven case study cities. Below, we refer to *CP, RP,* and equivalently Δ, M, which can be *[X]public*, or *[X]total*, if community adaptation and resilience are added/subtracted accordingly.

**2.3  Model Parameterization**

Service deficit (Δ) and service management (M) result from the interaction of the four (five) capitals. Service deficit represents the deficit in services at the citizen scale comprising deficits in access, continuity, reliability, affordability and safety (of water quality and of access). It is the combination of CP and RP that contributes to these aspects of water supply services. Equivalently, the maintenance and recovery of services is marshaled through capital availability and robustness determined by the parameters described below.

Parameter b is the sum of two additive processes: demand growth and service degradation. Capital availability is required for keeping up with demand growth, and capital robustness is required for service maintenance. Therefore, the lack of capital prevents urban managers from keeping up with demand growth, and lack of robustness leads to service degradation. So b is expressed by:

$$b = (1-CP)+(1-RP) \qquad (3)$$

Δ is recovered through M with efficiency *a*, which is defined as

$$a = \sum C_i \sum R_i \qquad (4)$$

Higher capital availability and robustness results in more efficient recovery of services (robustness for capitals W, I, F and preparedness to deal with shocks for capitals P and A; for brevity summarized as "robustness").

Coupling parameter $c_1$ determines the impact of service deficit on service management. Higher robustness buffers the impact of service deficit on service management. Therefore, when robustness is lacking, the recovery of M is limited:

$$c_1 = 1-RP \qquad (5)$$

According to Klammler et al. (2018), parameter r is the ratio of depletion over replenishment rates, and corresponds to the maximum depletion rate. Depletion of M is highest, when capitals and robustness are low. Thus, depletion rate *r* corresponds to average lack of robustness and capitals:

$$r = 1-(CP+RP)/2 \qquad (6)$$

Coupling parameter $c_2$ indicates the direct impact of shocks on service management. The ability to absorb shocks diminishes with diminishing capital availability and robustness. Therefore, the direct impact of shocks on service management is:

$$c_2 = r = 1-(CP+RP)/2 \qquad (7)$$

The scaling constant β signifies the scale at which degradation of M begins to level off.

$$\beta = RP \qquad (8)$$

i.e., when the level of robustness is reached.

The unitless coefficient *n* determines the steepness of the switch in service management as M reaches β, where higher *n*-values result in a steeper switch around β, while smaller *n*-values result in a more linear leveling off of service management degradation. *n* indicates how fast shocks impact M. It is set to

$$n = \sum R_i \qquad (9)$$

In the parameterization proposed above key model parameters are strongly correlated and determined by CP and RP. We assume that in urban systems, by definition CP ≠ 0.

### 2.3.1 Disturbance Regime

Various types of chronic and acute shocks impact different urban water systems, depending on a city's socio-political, economic, geographic, and climatic environment. Twelve types of threats resulting from four groups of hazards, which have the potential of producing shocks to the urban water supply system have been proposed by Krueger et al. (2019). Examples of chronic shocks are land subsidence causing infrastructure damage and contamination of piped water, competition for water resources, and illegal tapping into water pipes. Acute shock examples include earthquakes and landslides, industrial spills, war, or drought. A complete list of hazards resulting in chronic and acute shocks is provided as *Supplementary Information* (*SI)*.

The disturbance regime is characterized by the combination of chronic and acute shock time series. The number of shocks follow a Poisson distribution of mean frequency (density) $\lambda$ [1/T]; mean magnitude $\alpha$ [-] is drawn from an exponential distribution, with shock magnitudes relative to demand. Mean frequency of chronic shocks is:

$$\lambda_{chronic} = \frac{chronic\ shock\ score}{\sum chronic\ shocks}\ (1 + RP_{public})^{-1} \qquad (10)$$

where the shock score results from the summed binary scores (potential of occurrence = 1, exclusion of occurrence potential = 0) for each risk type divided by the sum of total potential risks. Adjustment by $RP_{public}$ indicates a city's ability to buffer shocks: According to Rodriguez-Iturbe et al. (1999), censoring (buffering) of shocks does not change shock magnitude, but results in a lower frequency of shocks. We apply this logic as suggested by Klammler et al. (2018) by censoring shocks proportional to $RP_{public}$. Acute shock frequencies are assumed to occur an order of magnitude less frequently:

$$\lambda_{acute} = \frac{acute\ shock\ score}{\sum acute\ shocks * 10}\ (1 + RP_{public})^{-1} \qquad (11)$$

Combined risks resulting from various causes can lead to supply intermittence or other disruptions in water services. Cities prepare for chronic shocks by installing isolation valves in the distribution networks to limit the affected population (Ozger & Mays, 2004). In case of a lack of adequate isolation valves within the network, entire distribution zones can be affected (Zischg et al., 2019). Distribution zone size depends on topography, network design, and operational strategies, and in the cases investigated here are in the order of 2-3% of the population (Abu Amra et al., 2011; CONAGUA, 2016). Acute shocks can affect large parts of the population. For example, the 2017 Central Mexico earthquakes left 25% of the population without water (Audefroy, 2018) and the 2003/2004 drought in Chennai, affected around 65% of the population (Srinivasan, 2008). Thus, we used mean magnitudes $\alpha_{chronic}$=0.03 for chronic, and $\alpha_{acute}$=0.2 for acute shocks.

We assessed the mean and maximum affected population based on actual isolation zones using isolation valve data for three cities. For Amman (data courtesy of Miyahuna Jordan Water Company), mean affected population (based on demand) was between 0.3-1.0%, while maximum affected population was between 6-38% for six distribution zones. Data analyzed for Ottawa, Canada (Jun, 2005) and Innsbruck, Austria, indicated maximum affected population of less than 1%. Differences between $\alpha_{chronic}$ and the stated values for affected population by isolation zones indicates the buffering capacity of different city types, which manifests in our model as a reduction in mean shock frequency (see Eq. 10 and 11).

Shock regimes are produced stochastically as the sum of time series of chronic and acute shocks, respectively, representing realistic scenarios of disturbances impacting water supply services. Shocks are added (subtracted) to service deficit and service management in each time step (see Eqs. 1 and 2).

### 2.4 Case Studies

We assess the resilience of urban water supply systems in seven cities on four continents. Cities were selected based on their contrasting water systems, which result from differences in capital availability, and lead to variability in water supply security. Three cities have fully developed capital portfolios and high levels of water security: Melbourne (Australia), Berlin (Germany) and Singapore. Following a drought that lasted more than a decade, Melbourne developed water infrastructure (large reservoir storage, desalination plants), that provides excess capacity during normal years, and is maintained at high financial cost (Ferguson et al., 2014). Berlin maintains availability of some excess water resources, which is a result of a decline in industrial production following the German reunification, and reduced domestic demand due to demand management measures (Moeller & Burgschweiger, 2008). Singapore maintains sufficient resources in a delicate balance of natural availability, water recycling, and water imports (Lee, 2005; Public Utilities Board (PUB), 2017; Ziegler et al., 2014).

Two cities have intermediate levels of CP: Amman (Jordan) and Mexico City (Mexico). Amman represents a city in transition aspiring water security, in spite of the country's water scarcity (water availability < 150 m$^3$cap$^{-1}$ y$^{-1}$). Urban water security is a key priority in this arid country (MWI, 2015). Large-scale investments (International Resources Group (IRG), 2013) and international agreements on transboundary water transfers (Klassert et al., 2015; Rosenberg et al., 2008), and continued support from international organizations and donors have allowed connecting close to 100% of the urban population to the piped network (Bonn, 2013; Rosenberg et al., 2007). Community adaptation mainly is a response to rationed water supply, forcing citizens to store water in rooftop tanks to bridge supply gaps during water supply intermittence (Rosenberg et al., 2007). High water abundance and low availability of all other capitals in Mexico City result in a transition water system that is characterized by the degradation of large-scale, inflexible water infrastructure, and water managers overwhelmed by ceaseless population growth, and large, frequent disruptions such as earthquakes (Lankao & Parsons, 2010; Tellman et al., 2018) with diverse strategies of community adaptation, including rooftop storage and access to private water vendors (Eakin et al., 2016).

Two cities have low CP: Chennai (India), where the lack of capitals for public services (*CP$_{public}$*) is only balanced by community adaptation (A) including self-supply from private wells and the private water market (Srinivasan et al., 2010). In Ulaanbaatar (Mongolia) deficits of all five capitals result in desperately low levels of services (Gawel et al., 2013; Myagmarsuren et al., 2015). Large inequality of water services in Ulaanbaatar is reflected in the operation of a split water system. The modern, central apartment areas receive piped warm and cold water at the household level, while around 60% of citizens live in the sprawling Ger areas (settlements lacking adequate infrastructure services) without access to piped water supply, sanitation, or roads. Ger residents have an average per capita water use of 8 lpcd, which they collect from water kiosks (Myagmarsuren et al., 2015). Frequent service interruptions due to frozen pipes are not uncommon, with temperatures as low as -40˚ Celsius. Ger residents drill shallow wells to access water directly on their property, and open-pit latrines substitute as sanitary infrastructure, which threatens the safety of the city's water sources (Myagmarsuren et al., 2015).

Cities faced with water service deficits often ration supply schedules, requiring citizens to store and treat water at the household level, and to supplement supplies through private services (Eakin et al., 2016; Klassert et al., 2015; Potter et al., 2010; Srinivasan et al., 2010). Intermittent supply through leaking pipes and dependence on shallow wells in proximity to dug latrines are significant water quality concerns (Gerlach & Franceys, 2009; Roozbahani et al., 2013; Sigel et al., 2012). Reliance on water sharing among households is yet another coping/adaptation strategy among households in poorer districts (Potter et al., 2010).

CPA data for the seven cities are provided in the *SI*. For details of the CPA analysis and more complete descriptions of these seven cities, see (Krueger et al., 2019).

## 3   Results

The model parameter input values resulting from the translation of the CPA for each of the seven cities are shown in **Table 1a)**. We use one or two parameterizations for each city: one for public services only, and, where public services do not meet demand, one for total services, which includes community adaptation of private households. In Ulaanbaatar, we also separately assess apartment (UB Apart) and Ger (UB Ger) areas in order to reveal the large inequality underlying the city's average service levels.

The model is solved for the following variables: 1) Fixed points of M and Δ (M$_{fix}$ and Δ$_{fix}$), which are stable points for public and total service deficit in each case study, respectively. Stable points are system attractors, towards which systems converge in the absence of shocks. 2) Mean values ($\mu$) and the coefficient of variation (CV) of M and Δ over the entire time series. 3) Crossing times (CT) are mean crossing times below and above a threshold defined by the expected mean values (M$_{thresh}$= M$_{fix}$ - c$_2$* $\alpha_{chronic}$; Δ$_{thresh}$= Δ$_{fix}$+ $\alpha_{chronic}$), which are a measure of the rapidity of service recovery after shocks. Numerical model results are presented in **Table 1b)**, as well as in **Figure 2** (additional figures are presented in the *SI)*.

**Table 1a:** Input parameters of systems dynamics model.

|  |  | Melbourne | Berlin | Singapore | Amman | Mexico City | Chennai | Ulaan-baatar | UB Apart | UB Ger | Amman | Mexico City | Chennai | Ulaan-baatar | UB Apart | UB Ger |
|---|---|---|---|---|---|---|---|---|---|---|---|---|---|---|---|---|
|  |  | public services | | | | | | | | | total services | | | | | |
| Model parameters | $r$ | -0.09 (0.01*) | 0.06 | 0.12 | 0.48 | 0.58 | 0.69 | 0.64 | 0.51 | 0.77 | 0.26 | 0.38 | 0.34 | 0.51 | 0.45 | 0.55 |
|  | $b$ | -0.18 (0*) | 0.12 | 0.24 | 0.96 | 1.15 | 1.38 | 1.28 | 1.03 | 1.54 | 0.53 | 0.77 | 0.67 | 1.01 | 0.91 | 1.11 |
|  | $n$ | 3.75 | 3.36 | 3.37 | 2.13 | 1.88 | 1.46 | 1.8 | 1.8 | 1.8 | 2.84 | 2.59 | 2.17 | 2.09 | 2.23 | 2.09 |
|  | $\beta$ | 0.94 | 0.84 | 0.84 | 0.53 | 0.47 | 0.36 | 0.45 | 0.45 | 0.45 | 0.71 | 0.65 | 0.54 | 0.52 | 0.56 | 0.52 |
|  | $a$ | 23.68 | 13.97 | 12.77 | 4.99 | 4.10 | 1.70 | 2.28 | 5.18 | 0.59 | 7.37 | 6.2 | 3.69 | 3.05 | 6.68 | 1.43 |
|  | $c_1$ | 0.06 | 0.16 | 0.16 | 0.47 | 0.53 | 0.64 | 0.55 | 0.55 | 0.55 | 0.29 | 0.35 | 0.46 | 0.48 | 0.44 | 0.48 |
|  | $c_2$ | -0.09 (0*) | 0.06 | 0.12 | 0.48 | 0.58 | 0.69 | 0.64 | 0.51 | 0.77 | 0.26 | 0.38 | 0.34 | 0.51 | 0.45 | 0.55 |
|  | $\lambda_{chronic}$ | 0 | 0.09 | 0.09 | 0.44 | 0.57 | 0.61 | 0.46 | 0.46 | 0.46 | *shocks same as for public services* | | | | | |
|  | $\lambda_{acute}$ | 0.02 | 0.01 | 0.02 | 0.03 | 0.05 | 0.05 | 0.05 | 0.05 | 0.05 | | | | | | |

\* Adjusted negative rate parameters *r* and *b*, and values for shock impacts ($c_2$).

**Table 1b:** Numerical results for urban case studies.

|  |  | Melbourne | Berlin | Singapore | Amman | Mexico City | Chennai | Ulaan-baatar | UB Apart | UB Ger | Amman | Mexico City | Chennai | Ulaan-baatar | UB Apart | UB Ger |
|---|---|---|---|---|---|---|---|---|---|---|---|---|---|---|---|---|
|  |  | public services | | | | | | | | | total services | | | | | |
| Numerical solutions | $M_{fix}$ | 0.99 | 0.96 | 0.92 | 0.48 | 0.25 | 0.01 | 0.19 | 0.29 | 0.08 | 0.81 | 0.69 | 0.65 | 0.4 | 0.56 | 0.31 |
|  | $\Delta_{fix}$ | 0 | 0.01 | 0.02 | 0.28 | 0.53 | 0.99 | 0.75 | 0.4 | 0.97 | 0.08 | 0.15 | 0.22 | 0.45 | 0.20 | 0.71 |
|  | $\mu_M$ | 0.99 | 0.96 | 0.92 | 0.46 | 0.17 | - | 0.14 | 0.24 | 0.07 | 0.80 | 0.67 | 0.63 | 0.36 | 0.54 | 0.28 |
|  | $\mu_\Delta$ | 0 | 0.01 | 0.02 | 0.30 | 0.65 | - | 0.81 | 0.47 | 0.98 | 0.09 | 0.16 | 0.23 | 0.49 | 0.21 | 0.75 |
|  | $CV_M$ [%] | 0 | 0.13 | 0.42 | 5.80 | 34.42 | - | 29.20 | 20.86 | 13.85 | 1.22 | 3.02 | 4.71 | 14.33 | 6.18 | 9.68 |
|  | $CV_\Delta$ [%] | $\infty$ | 60.82 | 45.62 | 9.12 | 15.31 | - | 7.06 | 0.27 | 0.78 | 16.13 | 13.12 | 14.76 | 4.26 | 15.12 | 4.43 |
|  | $CT_{Mbelow}$ | 0 | 0.90 | 0.90 | 1.16 | 1.62 | - | 1.46 | 1.46 | - | 1.16 | 1.62 | 1.68 | 1.46 | 1.46 | 1.46 |
|  | $CT_{\Delta above}$ | 0 | 0.04 | 0.05 | 0.19 | 0.32 | - | 0.44 | 0.26 | - | 0.10 | 0.13 | 0.22 | 0.30 | 0.14 | 0.45 |
|  | % failure | 0 | 0 | 0 | 26.7 | 100 | 100 | 100 | 97.2 | 100 | 0 | 0.8 | 1.0 | 64.5 | 9.1 | 95.1 |

For cities with a deficit in water services, community adaptation significantly improves urban water services, so that total services are much larger than public services. Due to the impacts of the shock regimes in all but the first three cities, mean values are significantly lower and higher than fixed points $M_{fix}$ and $\Delta_{fix}$, respectively. Mean, CT and CV are relatively low (see results for each city below). While higher values might be expected, the implementation of urban infrastructure serves the objective of reducing natural variability, which explains the relatively low variability of services compared to unmanaged systems, such as, e.g., river discharge. $CT_{\Delta above} < CT_{Mbelow}$ indicates that services are recovered faster than service management. $CV_\Delta$ increases with increasing services as a result of decreasing mean service deficit ($\mu_\Delta$).

The dynamic behavior of each system is contingent on the specific (stochastic) shock regime. We tested the probability of failure in response to shocks using a Monte Carlo approach, by running 1000 simulations x 1000 time units for each model parameterization. "Collapse" or "failure" of the system refers to the breakdown of services (public and/or total), for which the condition is $\Delta(t)=1$ and $M(t)=0$. The simulations terminate in the case of collapse. However, depending on the severity of the damage, the availability of support for reorganization and recovery, systems tend to be recovered after some lag time (see example for Chennai public services). Results are shown in the last row of Table 1b), as well as in **Figure 1**.

We plot the timing of collapse for all 1000 simulations in Fig. 1. The figure shows that, while failure probability is high for multiple case studies (see Table 1b), the "survival length" (time to collapse) is contingent on the realization of the shock regime. Outcomes of $t_{collapse}$ are highly variable across *and* within cities (compare UB Apart and UB Ger), varying over three orders of magnitude.

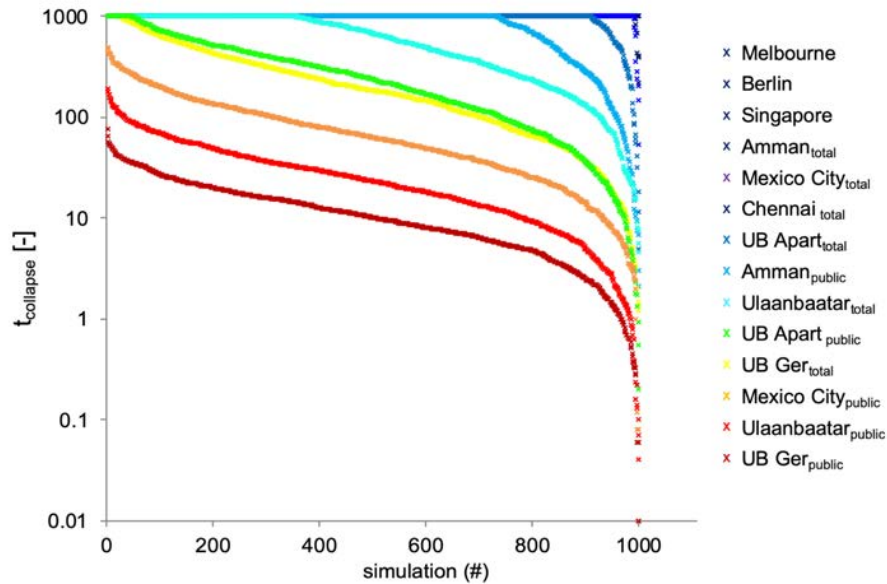

**Figure 1:** Time of collapse (in scaled time) for 1000 simulations for all case studies representing different realizations of stochastic shock regimes. No collapse was observed for Melbourne, Berlin, Singapore and Amman$_{total}$ (compare to Table 1b).

Based on the CPA assessment and model results, we present in the following three broad groups of cities. While there are no sharp boundaries between the groups and cities fall along a continuous gradient from low to high security and resilience, we use these broad categories for the convenience of organizing the presentation of results. 1) Water secure and resilient cities completely recover after shocks in all simulation runs; 2) Water insecure and non-resilient cities have a probability of failure ≈ 100 %; 3) Cities in transition have a failure probability significantly less than 100 %. Over time, cities in transition have managed to increase service security and resilience to a degree that mostly allows recovery from shocks, thereby escaping the poverty trap. They may also have transitioned into this "middle position" as a result of service decline (e.g., through high population growth, infrastructure degradation), where the inability to maintain high levels of security has led to a loss of services over time.

### 3.1 Water Secure and Resilient Cities

In this category, urban managers have reduced risks threatening their water security, so that both chronic and acute shocks occur at low frequencies (see **Table 1a**). Since the three cities represented here (Melbourne, Berlin, Singapore) also face relatively low rates of population growth, they are able to keep pace with demand growth and infrastructure degradation (small values of *b*). Potential shocks impacting services can be efficiently recovered thanks to high availability of capitals as well as system robustness (high efficiency *a*). Various factors, including high income levels of citizens, reliable accounting of water services, and anticipatory infrastructure maintenance (see case study descriptions in (Krueger et al., 2019)) keep depletion of service management low (rate constant *r*). System robustness in these cities leads to slow depletion of service management (large values for parameters *n* and β).

**Figure 2 a-c)** shows results for Melbourne, representative of water secure and resilient cities. Low magnitude, chronic shocks have been eliminated, causing no increase in Δ nor depletion of M. Acute shocks also have no impact on M. Even a large magnitude event occurring at time step 935 causes no impact on M, as the two systems, Δ and M, are decoupled (coupling parameters $c_1, c_2 ≈ 0$). Acute shocks impacting Δ are so quickly recovered that residents are unlikely to notice the deficit (t ≈ days or less). The decoupling between Δ and M indicates that urban managers have access to large amounts of capitals and robustness is high, so that any shock impacting their water systems can be buffered or quickly recovered without any impact on the ability of the managers to deal with recurring shocks. The degree of decoupling increases from Singapore, to Berlin, to Melbourne (decreasing values of $c_1$ and $c_2$) (see Table 1a).

The state-phase diagram (Fig. 2c) shows a single stable state with Δ ≈ 0 (no service deficit, i.e., full services) and M ≈ 1 (maximum service management capacity). The horizontal phase lines and the M-nullcline (horizontal at M=1) indicate that any magnitude event will only impact services, with response decoupled from M. Thus, under the assumption of static CP and RP, this type of city would always recover to a state with full services, even when faced with large shocks. In the case at hand, a 13-year drought hit Melbourne over the years 1997-2010, and urban managers invested in various infrastructure, including additional storage reservoirs, desalination plants, as well as adapting its governance system to be more responsive to droughts before critical reservoir level thresholds were crossed (Ferguson, Brown, et al., 2013). Thus, it did indeed buffer the drought and recover from the deficit in its reservoirs. Similarly, Singapore has (limited) buffer capacities for desalinating water when precipitation levels are insufficient to replenish the city's storage reservoirs (Ziegler et al., 2014). These results show dynamic behavior under current conditions, but do not make predictions or produce future scenarios, for which changing conditions (e.g., in RP) could be assumed. Such changes could produce multiple stable states, as illustrated in Figure 3. See Section 4 for further discussion.

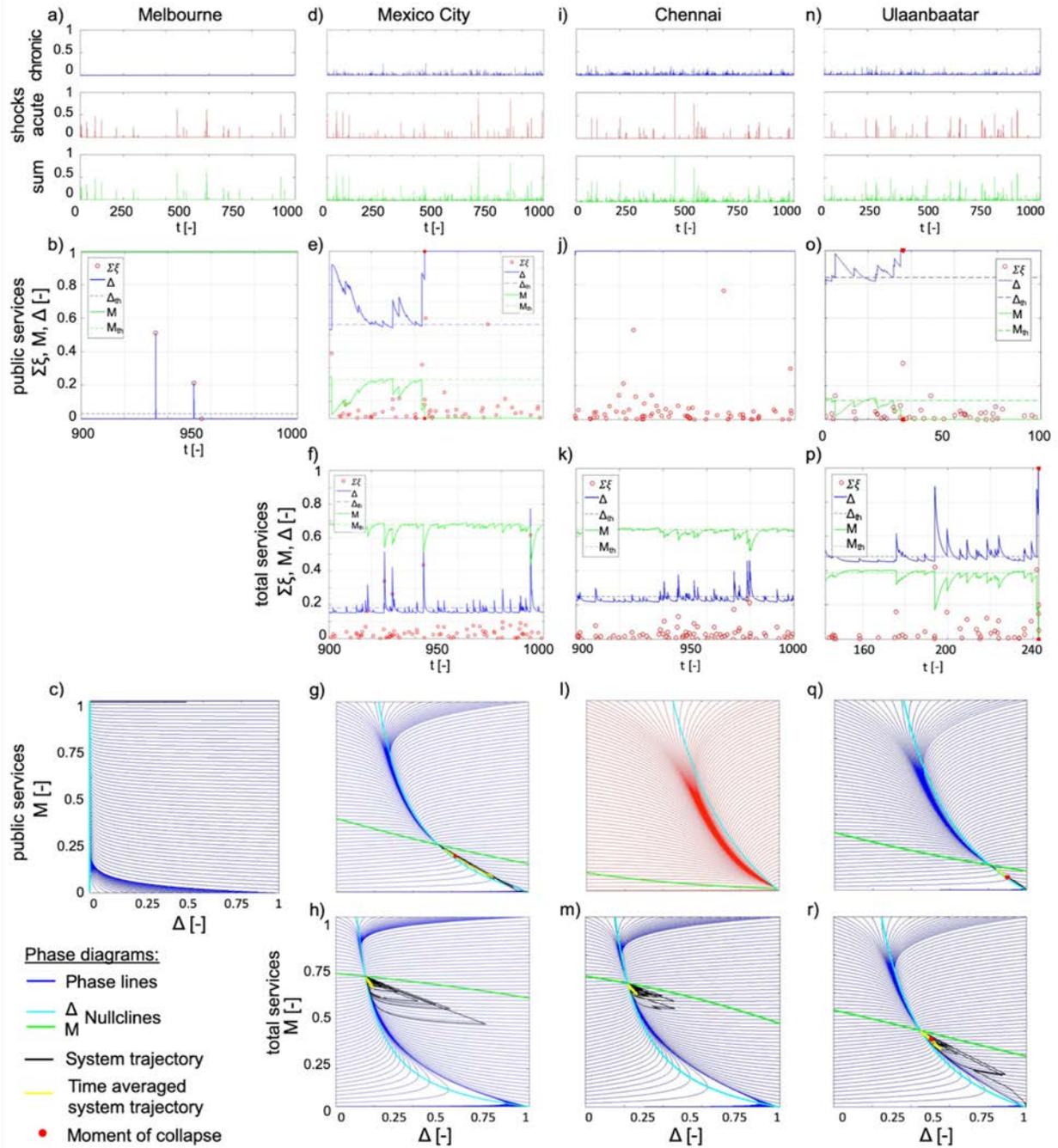

**Figure 2:** Time series and phase diagrams for different types of cities. Top panels: Time series of shocks (T=1000) with chronic (top row), acute (middle row) and combined shock regimes (bottom row). Center panels (rows 2 and 3): Time series of ξ, Δ and M and phase diagrams show trajectories for the last 100 time unit window, only, in order to better illustrate individual shock impact and recovery processes. Lower panels (rows 4 and 5): State-phase diagrams serve to identify stable states by running 100 model iterations for the phase trajectories (blue lines) to converge. Undisturbed phase trajectories (blue) converge toward a single stable point (intersection of Δ and M nullclines). System trajectories including shocks (black lines) correspond to time series of Δ and M. Yellow trajectories are time-averaged, red dot marks moment of collapse on time-averaged trajectory.

Panels **a-c)** show results for Melbourne as an example of resilient and water secure cities.

**Figure 2 continued:** Panels **d-h**) for Mexico City representing cities in transition. Shocks lead to system collapse of public services Panels g-h): Phase diagrams illustrate the increase in water security and resilience achieved through community adaptation compared to public services, only

Panels **i-r**) for Chennai and Ulaanbaatar illustrate system behavior for water insecure and non-resilient cities. Red phase lines for public services in Chennai (panel l) indicate convergence towards collapse in the absence of shocks (continuous degradation of services). Shocks lead to system collapse in Ulaanbaatar (panels o-r). Comparison of phase diagrams of public services (panels j, o and l, q) with total services (panels k, p and m, r) illustrate the regime change from low security and resilience towards an intermediate state, which is achieved through community adaptation. Time series for public services in Chennai are not produced (except shocks), as collapse occurs even in the absence of shocks (red phase lines). In the case of collapse (T<1000), the last 100 time units before collapse are shown (panels o, p).

### 3.2 Water Insecure and Non-Resilient Cities

In Ulaanbaatar (Mongolia) and Chennai (India) public water services cover only a fraction of demand due to a combination of water scarcity (economic or physical), lack of or decrepit infrastructure, as well as limited management capacity and rising demand due to population growth. The community adapts to insufficient services through various strategies, such as drilling private water wells, buying water on the private market (e.g., tanker trucks, stores), or using public facilities for laundry washing and personal hygiene, such as in public bath houses, or sharing water among neighbors (Myagmarsuren et al., 2015; Rosenberg et al., 2007; Sigel et al., 2012; Srinivasan et al., 2010).

**Figure 2 i-r)** summarizes the results for these two cities. Shocks occur frequently with both low and high magnitude (Fig. 2i,n), and they lead to collapse of both public as well as total services in Ulaanbaatar, as indicated by the red dots marking the moment of collapse in the time series in Figs. 2o) and 2p). Public services in Chennai are characterized by low CP and RP, and the city's water system thus converges towards collapse even in the absence of shocks, which is illustrated by the phase diagram in Fig. 2l), where red phase lines direct the system towards $\Delta \approx 1$ (no water services) and $M \approx 0$ (no capacity for service management), thus, a time series of shocks and recovery is not shown. In 2003/2004 a drought led to the complete suspension of piped water services (Srinivasan et al., 2013). However, the phase diagram for Chennai demonstrates that even in the absence of such shocks, population growth and increasing demand for water resources or competition among urban, peri-urban and agricultural sectors would ultimately lead the urban water system into collapse without additional investment into water security. Although public services in Ulaanbaatar are characterized by similarly low capital availability, slightly higher levels of robustness ($RP_{public}$=0.45 in Ulaanbaatar versus 0.36 in Chennai) keep the system just above collapse for the equilibrium solution (higher system robustness for Ulaanbaatar determines the dynamics of service management, Fig. 2q), compared to no dynamics in Chennai Fig. 2l).

Public services in these cities are indicative of a poverty trap, where inability to marshal the necessary capitals keeps systems precariously close to collapse (failure probability assessed in the Monte-Carlo simulation $\approx$ 100%, see Table 1b and Fig. 1). Community adaptation changes these cities' water security into a transitional state (Fig. 2m,r). The high adaptive capacity of Chennai's community (Srinivasan et al., 2013) leads to system resilience of total services, with recovery to relatively high levels of total services even from large shocks. Two consecutive large shocks to total services around t = 980 (Fig. 2k) or after around 80 years for unit t ≈ 1 month, represents an event such as the 2003/2004 drought, which the urban community was able to cope with through adaptive measures. For total services in Ulaanbaatar, large values of $c_2$ lead to a strong impact of shocks on service management, which is why recovery is not possible from large and recurring shocks, and system collapse occurs at t = 243 or after around 60 years for unit t ≈ 2-3 months (Fig. 2p). This response is aggravated due to non-linearity in the system model: for higher levels of Δ and M phase lines are horizontal, indicating that Δ is recovered first, before the recovery of M (area between two nullclines in lower right-hand corner of Fig. 2r) and closer to the green M-nullcline. The direction of the phase lines changes the closer the system moves towards the light blue Δ-nullcline, indicating that here service management is recovered first, as it is required to usher service deficit recovery. Where recovery from shocks is possible, a time lag resulting from strong coupling of service management to service deficit (large values of $c_1$), as well as low efficiency in recovery (low values of $a$) can be observed for both Ulaanbaatar and Chennai (Fig. 2k,p). This is in contrast to the immediate recovery for the resilient and water secure cities. Ulaanbaatar survived a shock at t=195 (after around 50 years), while a shock of the same magnitude at t=243 leads to collapse, because the system had not fully recovered from a shock that had occurred shortly before (t=241). This demonstrates the contingency of these urban water trajectories on specific shock scenarios.

Results for separate model runs for Ulaanbaatar's split water system for Ger and apartment areas for public and total services indicate that public services in Ger areas would not function without community adaptation, as residents are required to fetch their daily water needs from kiosks (Sigel et al., 2012) (see results in Table 1b); figures and data used for model parameterization are provided in the SI). Total services in Ger areas remain highly vulnerable, and recurring shocks make these services prone to collapse, in spite of improved community resilience (Fig. 1). Comparison of results between public services in Ulaanbaatar's Ger areas and public services in Chennai shows that in both cases service deficit is $\Delta_{public} \approx$ 100%. However, high capacity of Chennai's citizens to adapt to the highly deficient services results in a total service deficit of only $\mu_{\Delta total}$ =0.23, while it remains at $\mu_{\Delta total}$=0.75 in UB Ger (Table 1b).

Although apartment areas in Ulaanbaatar supposedly receive water continuously, we found that public services in these areas have an average deficit of around 40% without community adaptation, and around 20% with adaptation. Public service deficit results from low capacity of service management, and can result in collapse in response to recurring shocks due to slow recovery of services. High leakage rates are the result of a degraded distribution network in the central urban area, which dates back to the 1960's, and has not received any significant maintenance or replacement.

### 3.3 Cities in Transition

Capital availability for public services is significantly higher in Mexico City and Amman than in the non-resilient cities presented in Section 3.2 (see Table 1). In Mexico City, the combination of relatively frequent chronic and acute shocks, degraded system robustness (resulting in large values of $c_2$), which leads to significant impact on service management in response to shocks, as well as slow recovery after shocks due to strong coupling of Δ and M (large value of $c_1$) can cause public services to collapse (see Figs. 2e and 2g). Two large shocks occurring around t = 945 and t=995 (or after approximately 80 years for unit t ≈ 1 month) represent events, such as the September 2017 earthquakes. In Amman, all model parameters take intermediate values that describe this transitional status, including a reduction in the occurrence of shocks compared to the water insecure and non-resilient cities. Model results for service deficit in Amman are higher than observed values of public water supply, which covers around 76% of household demand in Amman (Krueger et al., 2019). However, this water is supplied on a rationed schedule with water delivered on 2.5 days per week on average, reducing the level of "service", and disturbances are frequent due to pipe bursts, in particular in house connections (data with courtesy of Miyahuna Jordan Water Company). The shock regime reflects the fact that shutting off or rescheduling delivery days become necessary for repairs and maintenance work, and network properties designed for continuous supply cause pressure variation within the pipe network. Amman's citizens access an additional 8-10% of their water demand from the private market (Klassert, pers. comm.), and have adapted to rationed water supply by storing water in rooftop and basement tanks. This increases total water services to around 80% (or 20% deficit) on average (see results in Table 1b), additional text and figures in the *SI*).

### 3.4 Resilience Landscape

To better understand the resilient behavior of urban water systems, we tested the entire parameter space by systematically varying CP and RP within the realistic range ($0 \leq CP \leq 1.4$; $0 \leq RP \leq 1$), which produces dependent changes to *b, r, β, $c_1$, $c_2$,* a* and *n*. We show service deficit as a function of CP and RP in **Figure 3.** Each point in Fig. 3 represents a fixed point. This three-dimensional surface represents a resilience landscape, as it shows all fixed points for the entire parameter space and indicates areas with possible bifurcation or regime shifts. Multiple fixed points appear in the parameter range CP>1 and RP < 0.3. This parameter range represents systems with excess capital availability and degraded robustness. Here, cities maintain excess capital at high cost in order to maintain security, but risk shifting into an alternate regime if robustness degrades below a threshold of RP < 0.3. Such a situation can be considered a rigidity trap (see discussion). While fixed points occur for the entire range of capital values (CP), no fixed points exist for RP<0.3, as long as CP<1.

Colored circles and dots in Fig. 3 represent the case studies with public and total services, respectively. Arrows indicate shock impacts: Arrow length represents maximum impact magnitude on M, and is a measure of the system's capacity to absorb shocks; arrow width is proportional to mean crossing times of service deficit above a specified threshold ($CT_{\Delta above}$; threshold = expected mean service deficit), and is a measure of the rapidity of service recovery after shocks. Cities are distributed within a confined area and along a gradient of decreasing service deficit, and increasing CP and RP resulting from the co-evolution of infrastructure and institutions (Padowski et al., 2016). Resilience, as indicated by the arrows, increases accordingly, with lower shock impacts and time to recovery with increasing CP and RP.

Visual inspection of Fig. 3 shows that CP, RP and resilience (indicated by the length and width or the arrows) are all positively related in the investigated cities and follow a somewhat linear trend (data correlation shown in *SI*). However, the fold in the resilience landscape for excess capital availability (CP>1) and low robustness (RP<0.3) demonstrates that the evolution of CP and RP can follow a non-linear path, and bears the risk of systems crossing a tipping point. We propose that systems in a state of (CP>1) and decreasing RP enter into a rigid regime. In this regime, security and resilience no longer co-evolve, and approach a tipping point that marks the boundary between high and low services (or collapse). This supports our first hypothesis on non-linearity and tipping points on a theoretical basis (although none of the case studies fall within this area).

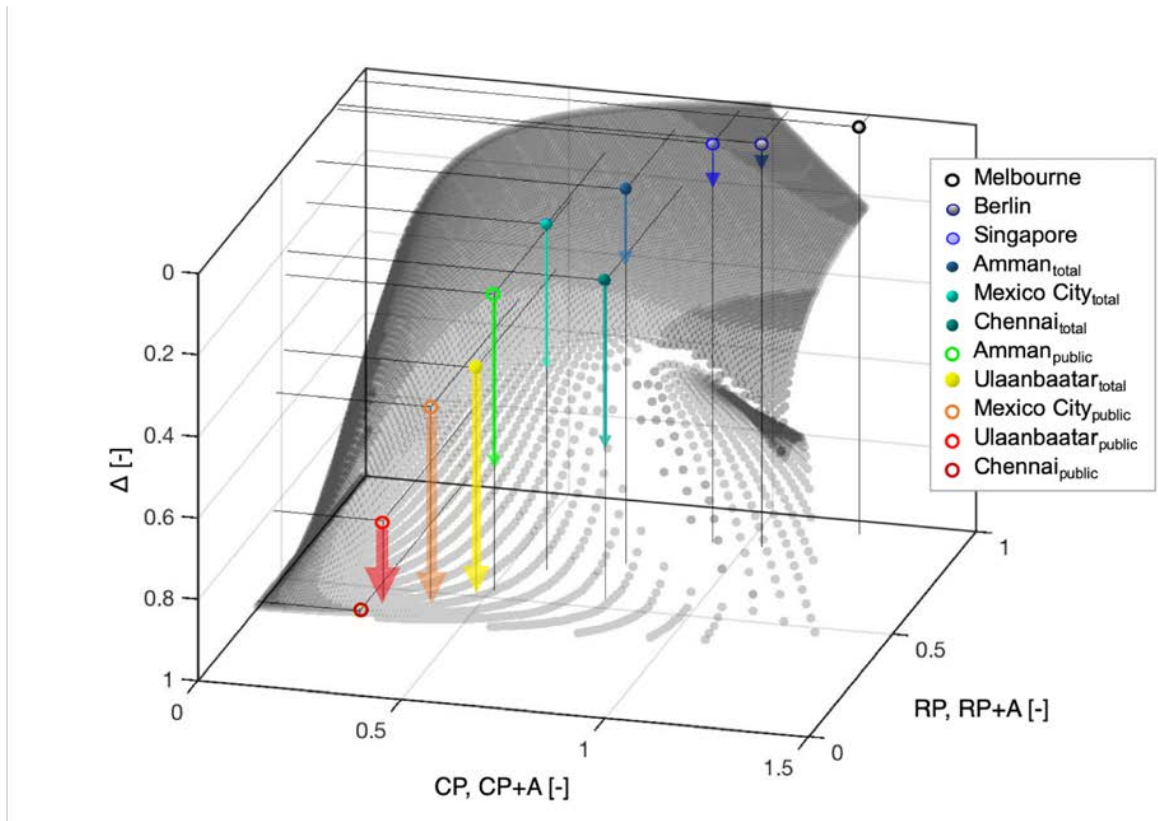

**Figure 3:** Resilience landscape simulated from systematic parameter variation across the entire parameter space and case study cities. Melbourne case study is positioned outside of the resilience landscape, because the model does not allow negative parameters, and we use adjusted parameters to calculate results for Δ (see Table 1a).
*a was calculated as a = (RP*4)*(CP*4).

In addition, we investigated changes in CP and RP (representative of urban water supply security), as well as (1-CT) as a proxy for the resilience of public versus total services. As can be seen from **Table 2**, the increase in these metrics (both absolute and relative) as a result of community adaptation is highly variable, which supports our second hypothesis. For example in Amman, the increases in resilience [(1-CT)+A] due to community adaptation is small (9% absolute and 12 % relative) compared to increases in security (CP+A=25 % and 50 %, RP+A 18 % and 34 % absolute and relative, respectively). In comparison, the increase in resilience is higher in Ulaanbaatar (14% absolute and 26% relative) compared to increases in security (CP+A=20 % and 73%, RP+A=7% and RP+A=16% absolute and relative, respectively).

**Table 2:** Effect of community adaptation on the security and resilience of urban water supply services (increases for total services due to community adaptation in comparison to public services).

| City | value (public services) | | | absolute increase (%) | | | relative increase (%) | | |
|---|---|---|---|---|---|---|---|---|---|
| | CP | RP | (1-CT) | CP+A | RP+A | (1-CT)+A | CP+A | RP+A | (1-CT)+A |
| Amman | 0.51 | 0.53 | 0.81 | 25.28 | 18.00 | 9.48 | 49.57 | 33.96 | 11.71 |
| Mexico City | 0.38 | 0.47 | 0.68 | 20.52 | 18.00 | 19.57 | 54.00 | 38.30 | 28.88 |
| Chennai | 0.25 | 0.36 | 0.00 | 53.44 | 18.00 | 78.46 | 213.78 | 50.00 | NA |
| Ulaanbaatar | 0.27 | 0.45 | 0.56 | 19.75 | 7.00 | 14.41 | 73.14 | 15.56 | 25.83 |

## 4    Discussion

Increasing pressures from global and climate change have launched a discussion about the adequacy of current water governance paradigms that seek to achieve or maintain water security by focusing on increasing supplies and managing water systems through command and control (Eakin et al., 2014; Kirchhoff et al., 2016; Larsen et al., 2016; Marlow et al., 2013; Varady et al., 2016). While the recent scholarship is strongly advocating new paradigms that embrace resilience thinking, promote adaptive capacity and favor more flexible, modular and sustainable strategies (Elmqvist et al., 2018, 2019; McPhearson et al., 2016; Meerow et al., 2016; Spiller et al., 2015; Webb et al., 2017), legacy effects of existing systems and slow uptake of such solutions explain why most urban water supply systems are still designed using conventional engineering solutions (Anderies et al., 2013; Marlow et al., 2013).

### 4.1    Along the Urban Water Security and Resilience Gradient

Here we show that under current conditions and governance paradigms, cities seeking to reliably provide water supply services are still able to co-develop the security and resilience of their water systems. Therefore, although the systems dynamics model used here has been shown to produce multiple stable states as expected for complex adaptive systems (Klammler et al., 2018), perhaps surprisingly, all represented urban case studies resulted in a single stable state (and collapse). Shocks and slow recovery can push systems away from their stable states even for extended periods of time. However, given the critical role of urban water services for the functioning of cities, our model results indicate that under current conditions, recovery is likely to the maximum stable state achievable given availability and robustness of capitals. Thus, cities fall along a continuous gradient on the urban water security and resilience landscape from water insecure and non-resilient to secure and resilient systems. Along this gradient, cities are distinguished according to their level of services and their response to shocks. We propose that, as cities grow and invest into their water supply systems, they evolve along the gradient from water insecure and non-resilient, to secure and resilient systems. Movement along such trajectories can occur as a transition from low to high security and resilience, and declining in the opposite direction.

Water insecure and non-resilient cities have low availability of capitals ($CP_{public} \lesssim 0.3$), leaving the majority of the population without adequate public services, and citizens are forced to turn to alternate services. Recovery from shocks is slow, and recurring shocks can quickly push such systems into collapse due to lack of robustness (here: Ulaanbaatar, Chennai). Increasing levels of CP and RP result in higher levels of services, but while in a transitional state (here: Amman and Mexico City), some level of service deficit remains and shocks continue to impact supply, requiring adaptive responses by the community.

The combined variability of public water services and community adaptive capacity result in large inequalities of total water services across cities, as well as within cities, and forces some communities to live with high service deficits. We showed that community adaptation is highly variable and constrained by, among others, the availability and access to alternative services. Adaptation is also subject to non-linear relationships of CP, RP and (1-CP). Place-based context provides insights into the meaning of quantitative values:

In Ulaanbaatar's Ger areas, harsh environmental and economic conditions limit community adaptation. In contrast, access to shallow groundwater through private wells and a private tanker market in Chennai allow the community to practically replace public services to cover its demand. Relatively reliable service management levels in Amman allow recovery from chronic and acute disturbances in spite of water scarcity. A high level of community adaptation to chronic disturbances (intermittence due to rationed supply schedules and frequent pipe bursts) improves continuity and reliability, rather than increasing volumetric water supply. Relatively low service management levels in Mexico City can result in collapse of public services in response to shocks. The history of Mexico City's urban water evolution demonstrates the legacy effect of decisions taken today or in the past on the long-term urban water trajectories, which can determine the evolution of entire cities (Bell & Hofmann, 2017; Marlow et al., 2013; Tellman et al., 2018).

Besides the reliance on private services, partly or in whole, such as in Amman, Chennai, Mexico City and Ulaanbaatar, hybrid systems (e.g., piped and trucked) also exist: Chennai's public utility hired private tanker trucks to deliver water bought from farmers' wells during the 2003/2004 drought (Ruet et al., 2007). Similarly, during the 2008 drought in Cyprus, water was delivered by tanker ships from mainland Greece (EEA, 2009), and in Legler (USA) water was delivered to citizens by truck after severe water contamination from a neighboring dump site (Edelstein, 2004, p. 55).

Averages are commonly used to represent entire cities, however intra-urban heterogeneity can be significant, with different urban water resilience regimes within the same city, as we demonstrated for Ulaanbaatar's split water system. In addition, although citizens adapt to insufficient services in order to meet their demands, both on a daily basis as well as in response to disasters (Béné et al., 2014; Brown & Westaway, 2011; Waters & Adger, 2017)) the relative cost incurred to these communities is disproportionate (Chelleri et al., 2015). While the system dynamics model demonstrated here does not quantify the cost of adaptation (economic, health, social, etc.), accounts of the local conditions illustrate the high price that communities pay for the little adaptation they can afford (Potter et al., 2010; Wutich & Ragsdale, 2008). Such conditions are characteristic of rapidly developing cities in many African and Asian countries, and concerns millions of people across the globe (Bell & Hofmann, 2017; Gopakumar, 2009; Sullivan, 2002; WORLD BANK, 2015; Zug & Graefe, 2014).

At the other end of the gradient, water secure and resilient cities have fully developed their capitals (CP ≈ 1), thus maintaining high levels of services and ability to instantly recover services even in the case of large and recurring shocks (here: Melbourne, Berlin and Singapore). Similar characteristics of the CPA, and thus model behavior are expected for cities in most of Europe, North America, Australia and parts of Asia.

While the development of excess water infrastructure to create a buffer against natural variability and the effects of drought, it also creates large sunk-cost effects and high maintenance costs, and can make cities inflexible in responding to changing demands and environmental conditions. Situations of excess capital such as those found in Melbourne apply to other economically advanced regions with large hydro-climatic variability, such as other cities in Australia and the US Southwest.

## 4.2 Possibility of Tipping Points

Our model-data analysis also shows the possibility of regime shifts after crossing a tipping point indicated by the fold in the resilience landscape for high levels of CP and low levels of RP. While we did not find data for cities existing in the rigidity trap, which marks the area of bi-stable states or tipping points, recent urban water (near-) emergency situations indicate increasing pressures resulting from global and climate change. This may change the trajectories of urban water systems with consequences on the co-evolution of urban water security and resilience, and thus, shift the predominance of cities residing in single stable state regimes toward multiple stable state regimes and into collapse. Emergency situations, which indicate such shifts are "Day Zero" scenarios in Cape Town in 2018 (Maxmen, 2018; Parks et al., 2019), threats of water rationing in Rome (Giuffrida & Taylor, 2017), and the recent drought hitting Chennai (India) (Jamwal, 2019). So far, returning rains or the installment of desalination plants in proximity to the coast (e.g., in the case of the millennium drought in Melbourne and other Australian cities) have allowed cities to recover to their original levels of services. However, changing rainfall patterns caused by climate change may permanently reduce the availability of water resources (IPCC, 2018), so that "excess infrastructure capital" (e.g., large storage reservoirs, river diversion projects or fossil-fueled, energy-intensive desalination plants) may no longer be a guarantee of urban water supply security. Therefore, gradual shifts in climate patterns, combined with the conventional response of enlarging the urban water footprint, could move systems into an area of the rigidity trap, where gradual loss of robustness and/or shocks can push cities across the tipping point into collapse. **Figure 4** schematically illustrates the point. Yellow arrows represent global change pressures pushing systems from resilient into rigid regimes.

Other factors can contribute to the loss of robustness, such as a growing global population, increasing competition or water quality impairments. Contamination of rivers has made potential sources of drinking water unusable in Beijing (China) (Qing, 2008; Tingting, 2017). Salinization of groundwater due to over-pumping is a concern in coastal and/or (semi-) arid areas around the world, such as in Amman (Hadadin et al., 2010), and the contamination of groundwater from agricultural and industrial pollution threatens the safety and sustainability of water supply, e.g., in Berlin (Henzler et al., 2014). Where current strategies for increasing security lead to excess capacity and "hides" the possibility of sudden service loss due to slow onset events or even in the absence of shocks ("false sense of security" (Ishtiaque et al., 2017)), such strategies may become decreasingly affordable under global change scenarios.

Global change pressures may also prevent cities from reaching security and resilience, as they evolve from low security and resilience through the transition phase. Current, supply-oriented strategies that focus on increasing supplies by enlarging urban water footprints may no longer satisfy urban water demands and push cities from transition into a rigid regime, instead of into a resilient regime (see Fig. 4a). Whenever possible, past experience and awareness of risk leads urban managers to develop capital robustness, hence creating the basis for resilient behavior. Thus, as long as cities maintain sufficient flexibility to adapt, social learning and adaptive management allow a co-evolution of security and resilience. However, internal factors (inflexibility/rigidity of developed systems) may also lead to lock-in, such as when cities build infrastructure to secure water resources, which they cannot afford to maintain over time, or which lose their reliability due to changing rainfall regimes as explained above. On the other hand, external factors (global change pressures) may constrain the "adaptation space". Thus, current management strategies imply an impending failure for all types of cities presented here.

Paradigm shifts towards demand management and closed water cycles will be needed in the future to achieve water security and resilience. First steps are being taken, such as developing a "water-sensitive city" in Melbourne (Ferguson, Frantzeskaki, et al., 2013) and efforts towards a closed urban water system in Singapore (Joo & Heng, 2017). However, as long as such strategies are developed with a sole focus on the water sector, trade-offs will remain that have the potential of constraining the long-term resilience of such systems (Lenouvel et al., 2014; Jianguo Liu et al., 2018; Paty Romero-Lankao et al., 2018)

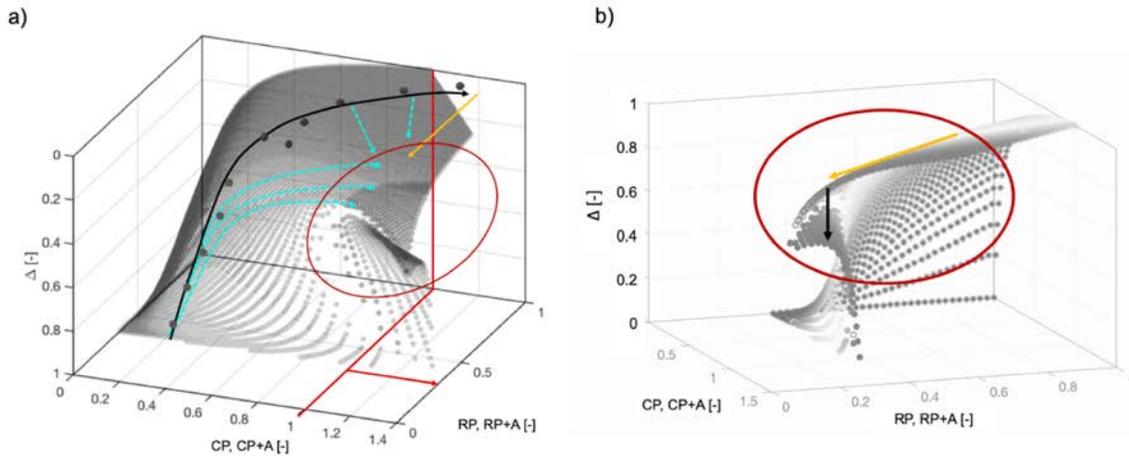

**Figure 4:** Schematic illustration of a tipping point in the rigidity trap with long-term trajectories of urban water systems under local and global change. Circled area indicates the rigidity trap. **a)** Proposed trajectory of water supply systems under current governance paradigms (black solid line with dots schematically representing case studies) and potential future pathways under local and global change pressures (light blue dashed lines). Local increase in capital availability attained through development of excess capital (red arrow pushing systems across the CP=1 threshold) and global change pressures (yellow arrow leading to a loss of RP) can push cities into the rigidity trap, illustrated by the fold in the resilience landscape (highlighted in Fig. 4b). Dashed arrows (light blue) pointing down indicate trajectories of "decline", dashed trajectory lines from bottom left indicate pathways of from low security and resilience into transition, constrained by global change pressures in achieving high security and resilience. **b)** Possibility of tipping points illustrated by the fold in the resilience landscape. Fig. 4b) is the same resilience landscape as in Fig. 3 and 4a), rotated around the vertical axis to highlight the bifurcation area.

### 4.3 Future Research

In spite of the aggregation of multiple capitals and robustness metrics, as well as several types of shocks into single variables, we were able to show dynamics of urban water systems, and outcomes are consistent with observations for the seven case study cities. Recovery times are fast when capital availability and robustness is high, while lack of CP and RP leads to slow recovery in non-resilient cities. Uncovering the accurate time scales will require comprehensive observational data of recovery times after a range of different types and magnitudes of shocks for multiple city types, which to date are rarely recorded (Cutter & Emrich, 2015).

Additional research and monitoring of relevant data are also necessary, in order to better understand the roles of each of the five capitals. Cascading and unexpected shock pathways could be accounted for in alternative shock terms, and application of a more refined disturbance concept would allow unpacking the interaction dynamics of CNHE systems in response to shocks (Grimm et al., 2017). Alternative models could be used to assess interactions between informal settlements and areas served by public services (e.g., tapping into public water pipes, movement of residents between water service areas for use of shower and laundry facilities in neighboring districts, etc.).

External drivers and social change resulting from global change processes will alter the balance between degradation and recovery, and the values of CP and RP, as well as shock regimes in the long-term. Changing the values of CP and RP over time and in response to shocks will allow modeling the long-term evolution of urban water security and resilience, with the resulting trajectories tracking the changing locations of fixed points of cities over time. Here, we present cities with different levels of water security development as a "space-for-time" concept, instead of reconstructing the long-term evolution of individual cities (see Fig. 4a). Future scenarios incorporating global change impacts may result in cities being pushed into the area of multiple stable states, where CP > 1 and RP < 0.3.

## 5   Conclusions

Our results expand the systematic understanding of urban water resilience with quantitative insights into the behavior of real-world CNHE systems in response to shocks and disturbances. This includes the emergence of stable states, resilience and regime shifts, as well as the role of community adaptation in the resilience of urban water supply systems. In particular, while we found that under current conditions, urban water systems tend to coevolve in terms of security and resilience, we propose that global change has the potential of driving systems across tipping points and alternate resilience regimes.

We found deep uncertainty in the resilience of urban water systems resulting from contingency on the disturbance regime, as indicated by the variability in survival periods for transition cities, and non-resilient and insecure cities. Whether or not a system can fully recover back to its stable state after a shock also depends not only on its capital portfolio, but also on the timing and magnitude of recurring shocks. This indicates that past experiences of shocks and recovery are not a valid indicator of future dynamics. Such uncertainty resulting from non-stationary forcing and temporal shifts in model parameters is problematic for predictions used for management. Thus, guidance provided should be in probabilistic terms, as with most models, not as deterministic forecasts. Compilation of data from several case studies, and incorporating long-term shifts in drivers of coupled systems dynamics are important directions for future research. Such data-model synthesis efforts are essential for developing guidance to urban managers and policy-makers for the future of urban water resilience.

We propose that the chosen case studies have archetypal character for representing cities with a comparable level of water system development. While the simulated time series cover intermediate scales of years to decades, we suggest that the long-term trajectories of cities over multiple decades to centuries is implied in a space-for-time principle, where the chosen case studies represent different levels of development and decline along the security and resilience gradient. The combined effect of increasing water scarcity and quality impairments of contested water resources will exacerbate water security issues. The possibility of tipping points indicated by the fold in the resilience landscape opens the question of sustainability. Scenarios of future urban water security demonstrate that global change will lead to changing patterns of resource availability, as well as increasing competition between sectors and cities (Floerke et al., 2018; Hoekstra & Mekonnen, 2012; Jenerette & Larsen, 2006). Therefore, future investigations should include sustainability considerations into their strategies for urban water security and resilience.

**Acknowledgments**

Research reported here was initiated in the frame of the Network Synthesis Workshop Series (2015-2018). The authors thank the workshop hosts and organizers, mentors and participants for fruitful discussions. The authors acknowledge the Jordanian Ministry of Water and Irrigation for assistance, and Miyahuna Jordan Water Company for sharing data used in this study. Financial support for EK and DB was provided by Helmholtz Centre for Environmental Research - UFZ, and for EK from Lynn Fellowship awarded by ESE-IGP, as well as from Purdue Climate Change Research Center (PCCRC) at Purdue University. PSCR financial support came from Lee A. Rieth Endowment in the Lyles School of Civil Engineering, Purdue University. This work was also supported by NSF Award No. 1441188. We thank the editor, Patricia Romero-Lankao, and two referees for valuable comments, which have greatly helped improve the manuscript. Data supporting the conclusions are listed in Tables 1 and 2, and in the SI.

**Table 3: List of Symbols and Abbreviations**

| | Description |
|---|---|
| A | Community Adaptation |
| a | efficiency constant of service recovery |
| b | growth rate of service deficit |
| C | Capital |
| $c_1$ | coupling parameter of service deficit onto service management |
| $c_2$ | coupling parameter: direct shock impact on service management |
| CNHE | coupled natural-human-engineered system |
| CP | capital portfolio |
| CPA | Capital Portfolio Approach |
| CT | mean crossing time |
| CV | coefficient of variation |
| F | Financial capital |
| I | Infrastructure |
| lpcd | liters per capita and day |
| M | service management |
| MC | Mexico City |
| n | coefficient |
| P | Management Power |
| r | maximum depletion rate of service management |
| R | Robustness |
| RP | Robustness Portfolio |
| t | time |
| UB | Ulaanbaatar |
| W | Water resources |
| $\alpha$ | mean shock magnitude |
| $\beta$ | scaling constant signifies scale at which M degradation begins to level off |
| $\Delta$ | service deficit |
| $\lambda$ | mean shock frequency |
| $\mu$ | mean value |
| $\xi$ | shocks |

Supporting Information for

# Resilience Dynamics of Urban Water Security and Potential of Tipping Points

E. H. Krueger[1,2], D. Borchardt[1], J. W. Jawitz[3], H. Klammler[4,5], S. Yang[2], J. Zischg[6] and P.S.C Rao[2,7]

[1]Department of Aquatic Systems Analysis, Helmholtz Centre for Environmental Research - UFZ, Leipzig, Germany
[2]Lyles School of Civil Engineering and Department of Agronomy, Purdue University, West Lafayette, Indiana, USA
[3]Soil and Water Sciences Department, University of Florida, Gainesville, Florida, USA
[4]Engineering School of Sustainable Infrastructure and Environment (ESSIE), University of Florida, Gainesville, Florida, USA
[5]Department of Geosciences, Federal University of Bahia, Salvador, Bahia, Brazil
[6]Unit of Environmental Engineering, Department for Infrastructure, University of Innsbruck, Innsbruck, Austria
[7] Department of Agronomy, Purdue University, West Lafayette, Indiana, USA

**Contents of this file**

**Introduction**

The Supplementary Information provided here contains input data used in the parameterization of the systems dynamics model, which is derived from the results of the CPA assessment for the seven case study cities (**Table S1**). Details of the analysis and input data to the CPA analysis can be found in Krueger et al. (2019). **Text S2** provides background information on the choice of the method for parameterization of the shocks used in the model. **Table S2** provides the corresponding input data, which was derived from Krueger et al. (2019). **Table S3** shows the metrics used for assessing the Capital Portfolio in Ulaanbaatar's apartment and Ger areas, respectively.

**Table S1.**     CPA results for Seven Case Study Cities

|  | $CP_{public}$ | $CP_{total}$ | $RP_{public}$ | $RP_{total}$ |
|---|---|---|---|---|
| **Melbourne** | 1.24 | | 0.94 | |
| **Berlin** | 1.04 | | 0.84 | |
| **Singapore** | 0.92 | | 0.84 | |
| **Amman** | 0.51 | 0.76 | 0.53 | 0.71 |
| **Mexico City** | 0.38 | 0.59 | 0.47 | 0.65 |
| **Chennai** | 0.25 | 0.78 | 0.36 | 0.54 |
| **Ulaanbaatar** | 0.27 | 0.47 | 0.45 | 0.52 |
| **UB Apart** | 0.52 | 0.64 | 0.45 | 0.56 |
| **UB Ger** | 0.01 | 0.37 | 0.45 | 0.52 |

**Table S1:** Assessment results from the Capital Portfolio Approach (CPA) for seven case study cities, and separate for apartment areas ("Apart") and Ger or slum areas ("Ger") for Ulaanbaatar (UB). Values are shown for public services ($CP_{public}$, $RP_{public}$) and total services ($CP_{total}$, $RP_{total}$), which includes community adaptation for accessing additional water resources, for storing and treating water at the household level, and for sharing water among neighbors, etc. Adapted from: (Krueger et al., 2019).

**Text S2.**     **Parameterization of Shocks**

While we use a Poisson process with exponentially distributed magnitudes for generating the shock terms, alternative shock distributions can be employed to represent specific known shock regimes. For example, earthquakes have been found to be power-law distributed both in terms of frequency and magnitude (Musson, Tsapanos, & Nakas, 2002). However, we maintain the same shock-generating probability distributions as in Klammler et al. (2018) for two reasons: 1) probability distributions of multiple different aggregated acute shock types, including earthquakes, economic crises or industrial spills, are assumed to occur randomly in the aggregate; and 2) data for time series of such shock regimes for the case studies that would prove a different distribution to be more appropriate are not available. To account for the fact that large events are less frequent, and noting that some shocks follow power-law, while other occur at random frequencies, we apply an exponential probability distribution for the magnitude of acute shocks, as well as frequencies following a Poisson distribution.

**Table S2.    Parameterization of Shocks**

| Risk category | Risk type description | shock type | Amman | Berlin | Chennai | Melbourne | Mexico City | Singapore | Ulaanbaatar |
|---|---|---|---|---|---|---|---|---|---|
| **Geological and geographic hazards** | earthquakes, tsunamis, volcanic eruptions, landslides | acute | 0 | 0 | 1 | 0 | 1 | 0 | 0 |
| | land subsidence | chronic | 0 | 0 | 1 | 0 | 1 | 0 | 1 |
| **Socio-economic and geo-political threats** | socio-economic/political changes/ unforeseen high immigration rates | chronic | 1 | 0 | 1 | 0 | 1 | 0 | 1 |
| | immediate threat of terrorism/war | acute | 0 | 0 | 0 | 0 | 0 | 0 | 0 |
| | competition for resources | chronic | 1 | 0 | 1 | 0 | 1 | 1 | 0 |
| | illegal tapping into water pipes | chronic | 1 | 0 | 1 | 0 | 1 | 0 | 1 |
| | immediate threat of economic crises | acute | 1 | 0 | 0 | 0 | 0 | 0 | 1 |
| **Contamination hazard** | industrial spills (upstream industry) | acute | 0 | 0 | 1 | 0 | 1 | 0 | 0 |
| | health impacts from degraded / lack of infrastructure | chronic | 1 | 0 | 1 | 0 | 1 | 0 | 1 |
| **Climate & weather-related hazards** | storms and wildfires (potential of damaging infrastructure) | acute | 0 | 0 | 1 | 1 | 1 | 1 | 1 |
| | floods/drought | acute | 1 | 1 | 1 | 1 | 1 | 1 | 1 |
| | extreme temperatures (freezing & bursting of pipes) | chronic | 0 | 1 | 0 | 0 | 0 | 0 | 1 |
| **Shocks** | mean risk frequency, $\lambda_{chronic}= \Sigma(score/6)*(1+RP)^{-1}$ | $\lambda_{chronic}$ | 0.44 | 0.09 | 0.61 | 0.00 | 0.57 | 0.09 | 0.46 |
| | $\lambda_{acute}= \Sigma(score/(6*10))*(1+RP)^{-1}$ | $\lambda_{acute}$ | 0.03 | 0.01 | 0.05 | 0.02 | 0.05 | 0.02 | 0.05 |

**Table S2**: Shock typology and occurrence probability in seven case study cities.

**Table S3: Metrics of CPA assessment for apartment and Ger areas in Ulaanbaatar**

|            | W    | I    | F    | P    | A    | CP   | CP+A | $A_R$ |
|------------|------|------|------|------|------|------|------|-------|
| Apartments | 1.19 | 0.78 | 0.66 | 0.25 | 0.12 | 0.41 | 0.54 | 0.43  |
| Gers       | 0.51 | 0.01 | 0.01 | 0.25 | 0.36 | 0.01 | 0.37 | 0.29  |

**Table S3:** Metrics of CPA assessment for apartment and Ger areas in Ulaanbaatar. Only values that differ from city average values are shown.

**Figure S1: Simulation results for water secure and resilient cities**

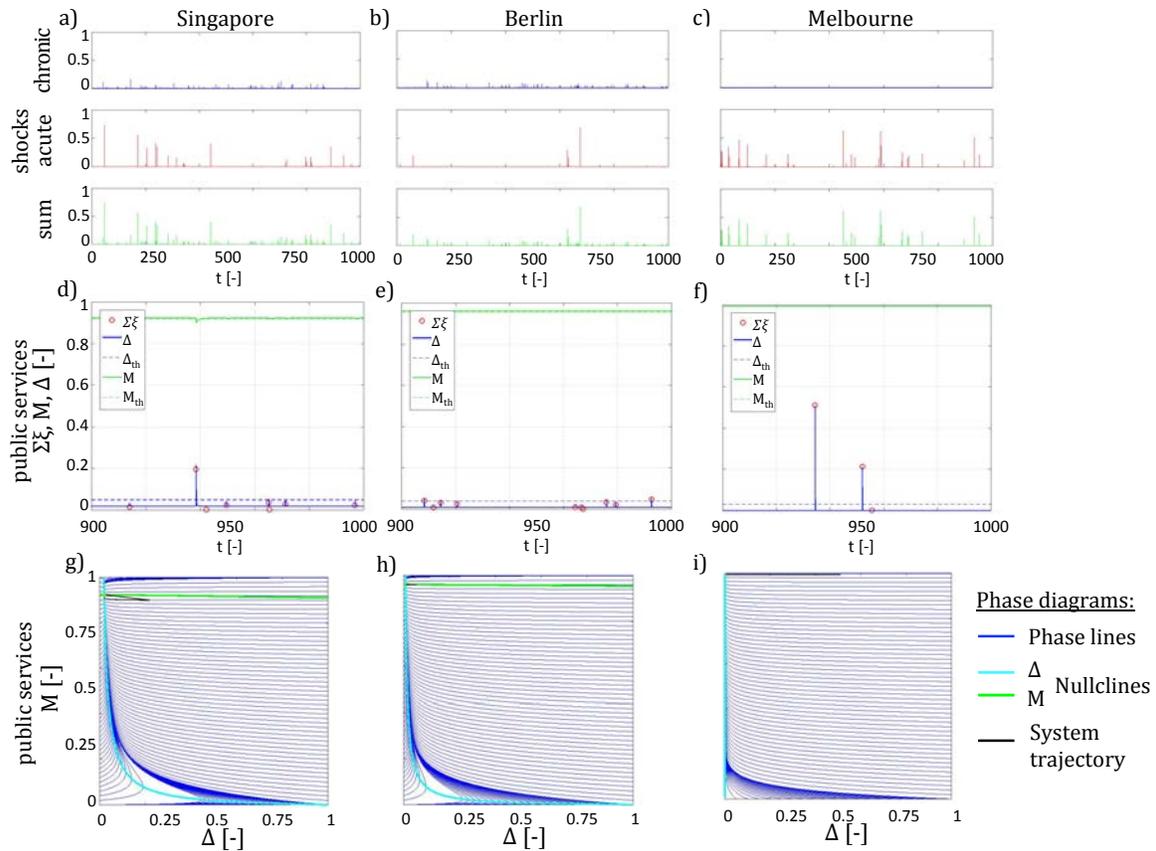

**Figure S1:** Upper panels: Time series of shocks (T=1000) with chronic (top row), acute (middle row) and combined shock regimes (bottom row).
Time series of $\Delta$ and M and phase diagrams show trajectories for a window of 100 time units, only (last 100 time units of total time series), in order to better illustrate individual shock impact and recovery processes. Center panels: Time series of $\xi$, $\Delta$ and M for 100 time units. Lower panels: State-phase diagrams serve to identify stable states by running 100 model iterations for the phase trajectories (blue lines) to converge. Undisturbed phase trajectories (blue) converge toward a single stable point (intersection of $\Delta$ and M nullclines). System trajectories including shocks (black lines) correspond to time series of $\Delta$ and M.

**Figure S2: Simulation results for water insecure and non-resilient cities**

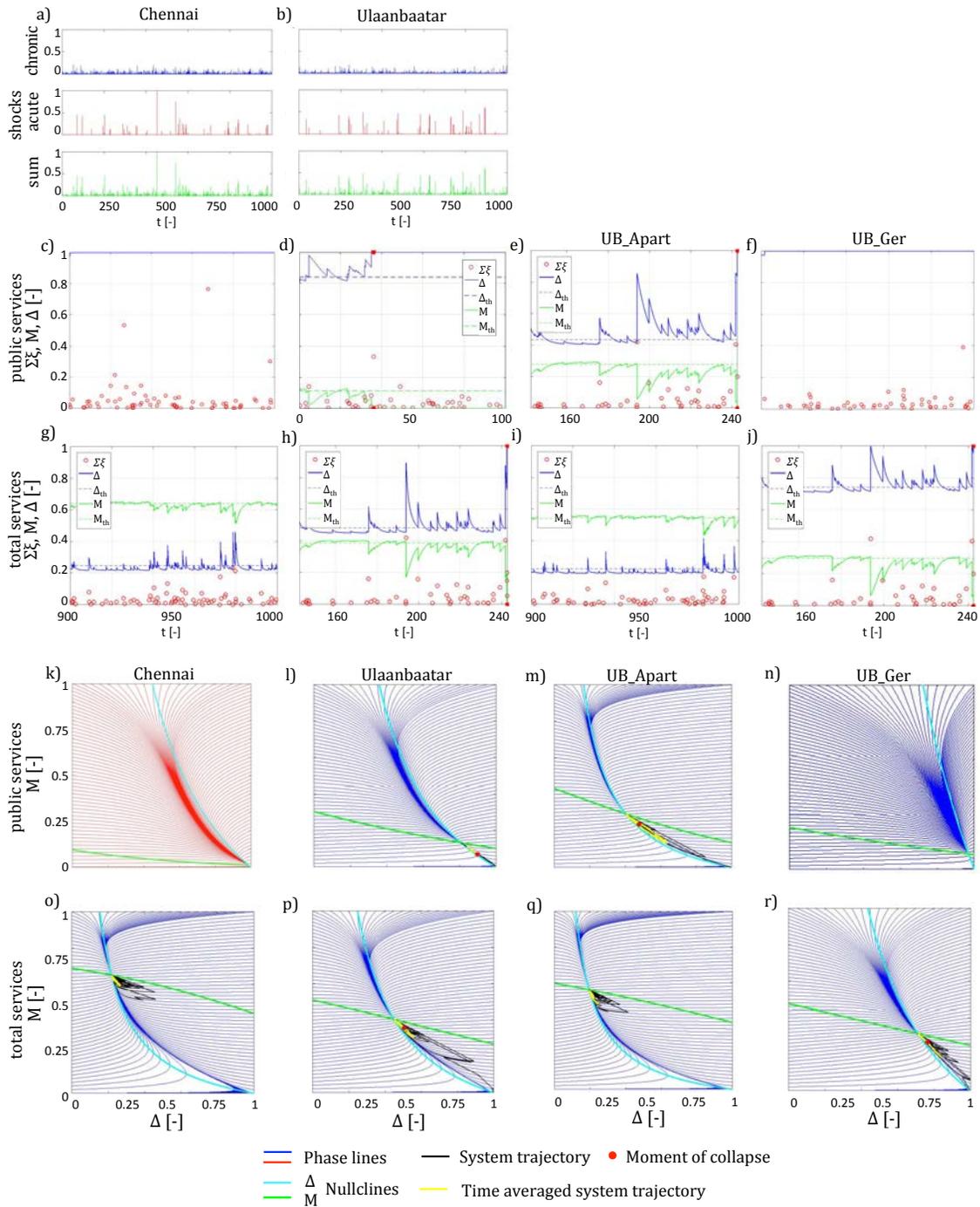

**Figure S2:** Time series and phase diagrams as in Fig. S1. Shocks lead to system collapse in Ulaanbaatar (panels d, h), while red phase lines for public services in Chennai indicate convergence towards collapse in the absence of shocks (continuous degradation of services). Comparison of phase diagrams of public services (panels k-l) with total services (panels o-p) illustrate the regime change from low security and resilience towards an intermediate state, which is achieved through community adaptation. Time series for public services in Chennai and in Ger areas are not produced (except shocks), as the first shock leads to collapse (UB_Ger), or collapse occurs even in the absence of shocks (red phase lines in panel k). In the case of collapse (T<1000), the last 100 time units before collapse are shown.

**Figure S2: Simulation results for water insecure and non-resilient cities**

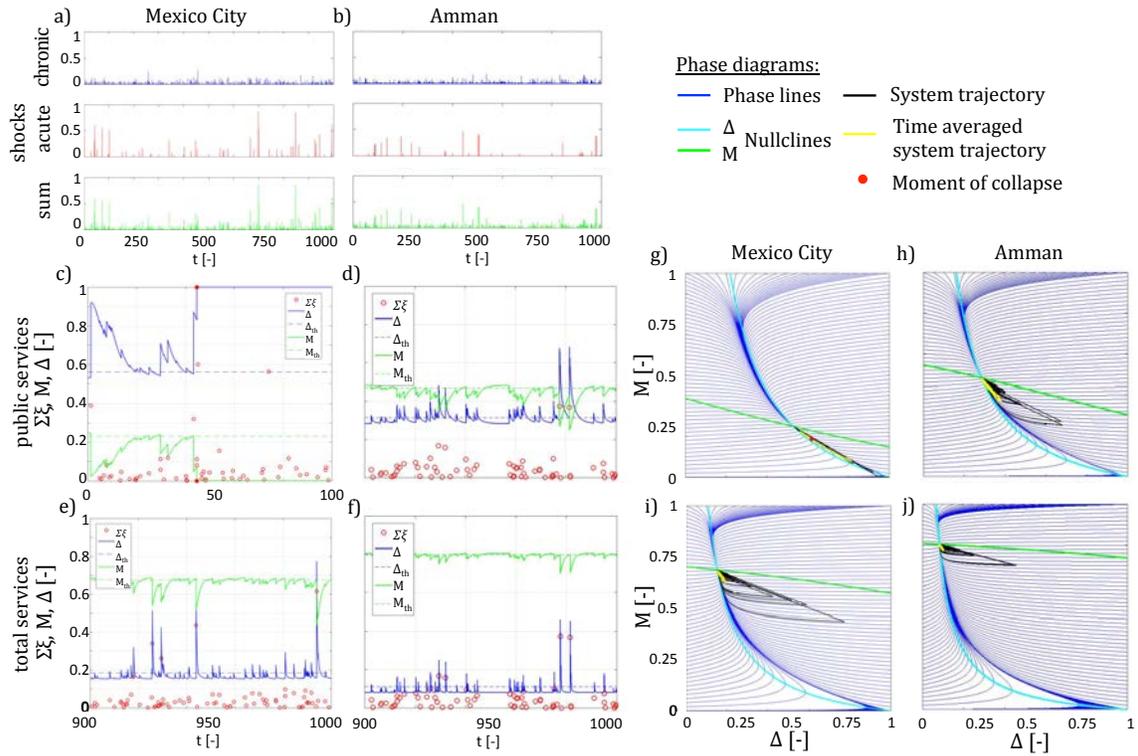

**Figure S3:** Cities in transition. Time series and phase diagrams as in Figs. S1 and S2. Shocks lead to system collapse of public services in Mexico City (panels c, g). Panels g-j): Phase diagrams illustrate the increase in water security and resilience achieved through community adaptation compared to public services, only. Amman has been receiving additional water through long distance water imports (Disi Conveyance Scheme) since 2013 (Miyahuna, 2014), which should be seen as increased services in the time series. However, water demand increased simultaneously as a result of population growth triggered by the Syrian refugee crisis (UNHCR, 2016), leveling out the additional supply. The shift in supply and demand could therefore have been noticed in the population as a short-term disturbance, such as occurring around t=980 or after around 40 years for unit t ≈ 2-3 weeks.

**Figure S4: Correlation between CP, RP and (1-CT)**

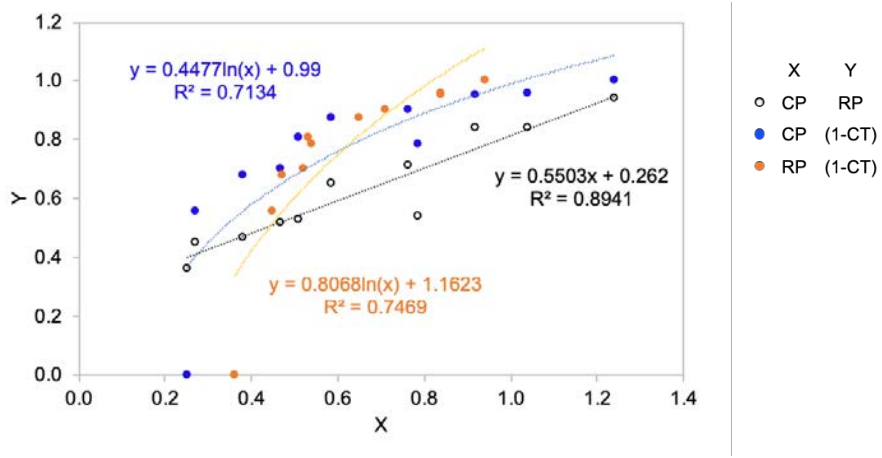

**Figure S4:** Correlations between CP, RP and (1-CT).